\documentclass[runningheads]{llncs}

\usepackage[T1]{fontenc}
\usepackage{graphicx}
\usepackage{amsmath}
\usepackage{wrapfig}
\usepackage{cite}
\usepackage[ruled, linesnumbered]{algorithm2e}
\usepackage{caption}
\usepackage{array}

\begin{document}

\title{Modality Cycles with Masked Conditional Diffusion for Unsupervised Anomaly Segmentation in MRI}

\titlerunning{Modality Cycles with Masked Conditional Diffusion}

\author{Ziyun Liang$^{1}$, Harry Anthony$^{*,1}$, Felix Wagner$^{*,1}$, Konstantinos Kamnitsas$^{1,2,3}$}

\authorrunning{Liang et al.}

\institute{$^1$Department of Engineering Science, University of Oxford, Oxford, UK
\\
$^2$Department of Computing, Imperial College London, London, UK \\
$^3$School of Computer Science, University of Birmingham, Birmingham, UK\\
\email{ziyun.liang@eng.ox.ac.uk}\\
* equal contribution}

\maketitle       
\begin{abstract}

Unsupervised anomaly segmentation aims to detect patterns that are distinct from any patterns processed during training, commonly called abnormal or out-of-distribution patterns, without providing any associated manual segmentations. Since anomalies during deployment can lead to model failure, detecting the anomaly can enhance the reliability of models, which is valuable in high-risk domains like medical imaging.
This paper introduces Masked Modality Cycles with Conditional Diffusion (MMCCD), a method that enables segmentation of anomalies across diverse patterns in multimodal MRI. The method is based on two fundamental ideas. First, we propose the use of cyclic modality translation as a mechanism for enabling abnormality detection. Image-translation models learn tissue-specific modality mappings, which are characteristic of tissue physiology. Thus, these learned mappings fail to translate tissues or image patterns that have never been encountered during training, and the error enables their segmentation.
Furthermore, we combine image translation with a masked conditional diffusion model, which attempts to `imagine' what tissue exists under a masked area, further exposing unknown patterns as the generative model fails to recreate them. We evaluate our method on a proxy task by training on healthy-looking slices of BraTS2021 multi-modality MRIs and testing on slices with tumors. We show that our method compares favorably to previous unsupervised approaches based on image reconstruction and denoising with autoencoders and diffusion models. Code is available at: \url{https://github.com/ZiyunLiang/MMCCD}

\end{abstract}

\section{Introduction}
Performance of deep learning-based image analysis models after deployment can degrade if the model encounters images with `anomalous' patterns, unlike any patterns processed during training. This is detrimental in high-risk applications such as pathology diagnosis where reliability is paramount. An approach attempting to alleviate this is by training a model to identify all possible abnormalities in a \textit{supervised} manner, using methods such as outlier exposure \cite{hendrycks_deep_2018,guha_roy_does_2022,tan_efficient_2021,tan_detecting_2022}. However, these methods are often impractical due to the high cost associated with data collection and labeling, in conjunction with the diversity of abnormal patterns that can occur during deployment. This diversity is enormous and impractical to cover with a labeled database or to model it explicitly for data synthesis, as it includes any irregularity that can be encountered in real-world deployment, such as irregular physical compositions, discrepancies caused by different acquisition scanners, image artifacts like blurring and distortion and more. Thus, it is advantageous to develop approaches that can address the wide range of abnormalities encountered in real-world scenarios in an \emph{unsupervised} manner.
Unsupervised anomaly segmentation works by training a model to learn the patterns exhibited in the training data. Consequently, any image patterns encountered after deployment that deviate from the patterns seen during training will be identified as \emph{anomalies}. In this paper, we delve into unsupervised anomaly segmentation with a focus on multi-modal MRI.

\noindent \textbf{Related Work:}
Reconstruction-based methods are commonly employed for unsupervised anomaly segmentation. During training, the model is trained to model the distribution of training data. During deployment, the anomalous regions that have never been seen during training will be hard to reconstruct, leading to high reconstruction errors that enable the segmentation of anomalies. This approach employs AutoEncoders (AE), such as for anomaly segmentation in brain CT \cite{sato_primitive_2018} and MRI \cite{atlason_unsupervised_2018, baur_deep_2019}, as well as Variational AutoEncoders (VAE) 
for brain CT\cite{pawlowski_unsupervised_2018} and MRI\cite{baur_deep_2019, shen_unsupervised_2019}.
AE/VAEs have attracted significant interest, though their theoretical properties may be suboptimal for anomaly segmentation, because the encoding and decoding functions they model are not guaranteed to be tissue-specific but may be generic, and consequently they may generalize to unknown patterns.
Because of this, they often reconstruct anomalous areas with fidelity which leads to false negatives in anomaly segmentation \cite{yan_learning_2021}. Moreover, AE/VAEs often face challenges reconstructing complex known patterns, leading to false positives \cite{saxena_generative_2023}. Generative Adversarial Networks (GANs)\cite{goodfellow_generative_2014} and Adversarially-trained AEs have also been investigated for anomaly segmentation, such as for Optical Coherence Tomography\cite{schlegl_f-anogan_2019} and brain MRI\cite{baur_deep_2019} respectively. However, adversarial methods are known for unstable training and unreliable results. 

A related and complementary approach is adding noise to the input image and training a model to denoise it, then reconstructing the input such that it follows the training distribution. 
At test time, when the model processes an unknown pattern, it attempts to `remove' the anomalous pattern to reconstruct the input so that it follows the training distribution, enabling anomaly segmentation. This has been performed using Denoising AutoEncoders (DAE)\cite{chen_deep_2018,kascenas_denoising_2022}. It has been shown, however, that choice of magnitude and type of noise largely defines the performance \cite{kascenas_denoising_2022}. 
The challenge lies in determining the optimal noise, which is likely to be specific to the type of anomalies. Since we often lack prior knowledge of the anomalies that might be encountered after model deployment, configuring the noise accordingly becomes a practical challenge. 
A related compelling approach is Denoising Diffusion Probabilistic Models (DDPM)\cite{ho_denoising_2020}, which iteratively reconstruct an image by reversing the diffusion process. \cite{pinaya_fast_2022} and \cite{bercea_mask_2023} employed \emph{unconditional} diffusion models by adding Gaussian noise to the input and denoising it, based on the assumption the noise will cover the anomaly and it will not be reconstructed. At the same time, both of them added masks to cover the anomaly. These unconditional Diffusion methods relate to DAE, as they also denoise and reconstruct the input image, but differ in that noise is added during test time and not only during training.

\begin{figure}
\includegraphics[width=1\textwidth]{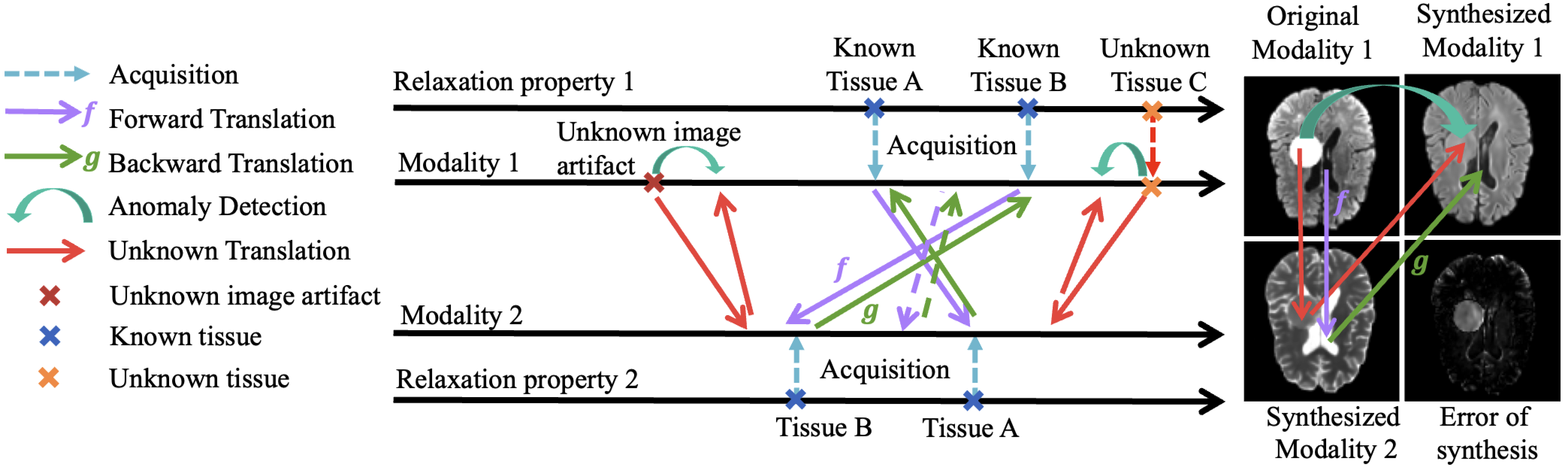}
\caption{Illustration of the multi-modal cyclic translation. MRI acquisition assigns different image intensities to different tissues. The cyclic translation model, shown as forward $f$ and backward $g$ translation respectively, learns the mappings between the two modalities for tissues within the training data. The mapping is specific to tissue physiology. Consequently, if unknown patterns are processed after deployment, the model will make random translations, leading to errors in the cyclic translation that indicate the presence of an anomaly. The right panel demonstrates how known tissues and anomalies are cyclically translated.}
\label{fig1}
\end{figure}

\noindent \textbf{Contributions:}
This paper introduces a novel unsupervised anomaly segmentation method for tasks where multi-modal data are available during training, such as multi-modal MRI. The method is based on two components.
First, we propose the use of cyclic-translation between modalities as a mechanism for enabling anomaly segmentation. An image translation model learns the mapping between tissue appearance in different modalities, which is characteristic of tissue physiology. As this mapping is tissue-specific, the model fails to translate anomalous patterns (e.g. unknown tissues) encountered during testing, and the error enables their segmentation. This differs from reconstruction- and denoising-based approaches, where the learned functions are not necessarily tissue-specific and may generalize to unseen patterns, leading to suboptimal anomaly segmentation. Furthermore, by employing cyclic-translation with two models, mapping back to the original modality, we enable inference on uni-modal data at test time (Fig.\ref{fig1}).
Additionally, we combine image-translation with a Conditional Diffusion model. By using input masking, this conditional generative model performs image translation while re-creating masked image regions such that they follow the training distribution, thus removing anomalous patterns, which facilitates their segmentation via higher translation error. We term this method Masked Multi-modality Cyclic Conditional Diffusion (MMCCD). In comparison to previous works using \emph{unconditional} Diffusion models with masking \cite{pinaya_fast_2022,bercea_mask_2023}, our work uses \emph{conditional} Diffusion, different masking strategy, and shows the approach is complementary to image-translation.
We evaluate our method on tumor segmentation using multi-modal brain MRIs of BraTS`21\cite{bakas_advancing_2017, baid_rsna-asnr-miccai_2021}, a proxy task often used in related literature. We train our model on healthy-looking MRI slices and treat tumors as anomalies for segmentation during testing. 
We demonstrate that simple Cyclic-Translation with basic Unet outperforms reconstruction and denoising-based methods, including a state-of-the-art Diffusion-based method. 
We also show that MMCCD, combining cyclic-translation with masked image generation, improves results further and outperforms all compared methods.

\section{Multi-modal Cycles with Masked Conditional Diffusion}

\subsubsection{Primer on multi-modal MR:} MRI (Magnetic Resonance Imaging) is acquired by exciting tissue hydrogen nuclei with Radio Frequency (RF) energy, followed by relaxation of the nuclei. While nuclei relax and return to their low-energy state, a RF signal is emitted. The signal is measured and displayed with the shades of gray that form an MR image. 
Nuclei of different tissues have different relaxation properties when excited by the same RF pulse. Consequently, the RF signal they emit during relaxation also differs, creating the contrast between shades of gray depicting each tissue in an MRI. 
However, certain tissues may demonstrate similar relaxation properties when excited with a certain RF pulse, thus the corresponding MRI may not show the contrast between them. Therefore, different MRI modalities are obtained by measuring the relaxation properties of tissues when excited by different RF pulses. As a result, in multi-modal MRI, a tissue is considered visually separable from any other when it exhibits a unique combination of relaxation properties under different excitation signals, i.e. a unique combination of intensities in different modalities.

\noindent \textbf{Cyclic Modality Translation for anomaly segmentation: } We assume the space of two MRI modalities for simplicity, $\mathcal{X}$ and $\mathcal{Y}$. When a model $f$ is trained to translate an image from one modality to another, $f: \mathcal{X} \rightarrow \mathcal{Y}$, it learns how the response to different RF pulses changes for a specific tissue, which is a unique, distinct characteristic of the tissue.
Based on the insight that the modality-translation function is highly complex and has a unique form for different tissue types, we propose using modality-translation for anomaly segmentation. This idea is based on the assumption that in the presence of an anomalous image pattern, such as an `unknown' tissue type never seen during training, or an out-of-distribution image artifact, the network will fail to translate its appearance to another modality, as it will not have learned the unique and multi-modal `signature' of this anomalous pattern.
Consequently, a disparity will be observed between the synthetic image (created by translation) and the real appearance of the tissue in that modality, marking it as an abnormality.

We assume training data consist of $N$ pairs of images from two modalities, denoted as $D_{tr}=\left\{\mathbf{x}^{i},\mathbf{y}^{i}\right\}^N_{i=1} $, where $\mathbf{x} \in \mathcal{X}$ and $\mathbf{y} \in \mathcal{Y}$. A mapping $f\!: \mathcal{X} \rightarrow \mathcal{Y}$ is learned with a translation network $f$. We denote as $\mathbf{\overline{y}} = f(\mathbf{x})$ the synthetic image in $\mathcal{Y}$ predicted by the translation network. Using the training image pairs $D_{tr}$, we train a model to learn the mappings between modalities.

During testing, the model processes unknown data $ D_{te}=\left\{\mathbf{x}^{i}\right\}^N_{i=1} $. The model may be required to process images with anomalous patterns. Such patterns can be unknown tissue types or image artifacts never seen during training (Fig.\ref{fig1}). Our goal is to detect any such anomalous parts of input images and obtain a segmentation mask that separates them from `normal' parts of input images, i.e. patterns that have been encountered during training.

We first perform the learned translation $f\!: \mathcal{X} \rightarrow \mathcal{Y}$, obtaining $\mathbf{\overline{y}}=f(\mathbf{x})$ for a test sample $\mathbf{x}\in D_{te}$. If we had the real image $\mathbf{y} \in \mathcal{Y}$, we could separate an anomaly by computing the translation error $\mathbf{a}$, which is the L2 distance between the translated image $\mathbf{\overline{y}}$ and $\mathbf{y}$: $\mathbf{a} = \lVert \mathbf{y}-\mathbf{\overline{y}} \rVert _2 > h$, where $h$ denotes the threshold for detecting the anomaly. As we assume $\mathbf{y} \in \mathcal{Y}$ is not available during training, 
we separately train another network $g:\mathcal{Y}\rightarrow\mathcal{X}$ to translate images from modality $\mathcal{Y}$ back to modality $\mathcal{X}$ as $\mathbf{\overline{x}} = g(\mathbf{\overline{y}})=g(f(\mathbf{x}))$. In this way, a cyclic translation is performed and only images from a single modality $D_{te}=\left\{\mathbf{x}^{i}\right\}^N_{i=1}$ are required for inference/testing. The networks will translate a test image $x$ from modality $\mathcal{X}$ to $\mathcal{Y}$, then back to $\mathcal{X}$ without requiring testing images from modality $\mathcal{Y}$:
\begin{equation}
\overline{\mathbf{x}} = g(f(\mathbf{x}))
\end{equation}
Then the anomaly is detected by computing the translation error:
\begin{equation}
\mathbf{a} = \lVert \mathbf{x}-\mathbf{\overline{x}}\rVert _2 > h
\end{equation}

The cyclic translation with the two models $f$ and $g$ is illustrated in Fig.\ref{fig1}. Our goal is to detect any anomalous pattern in input image $\mathbf{x}\in \mathcal{X}$ from the cyclic translation error $\mathbf{a}$. It is worth noting at this point that this cyclic mapping process can be modeled by any existing translation network like basic UNet\cite{ronneberger_u-net_2015}.

\begin{figure}
\includegraphics[width=0.8\textwidth]{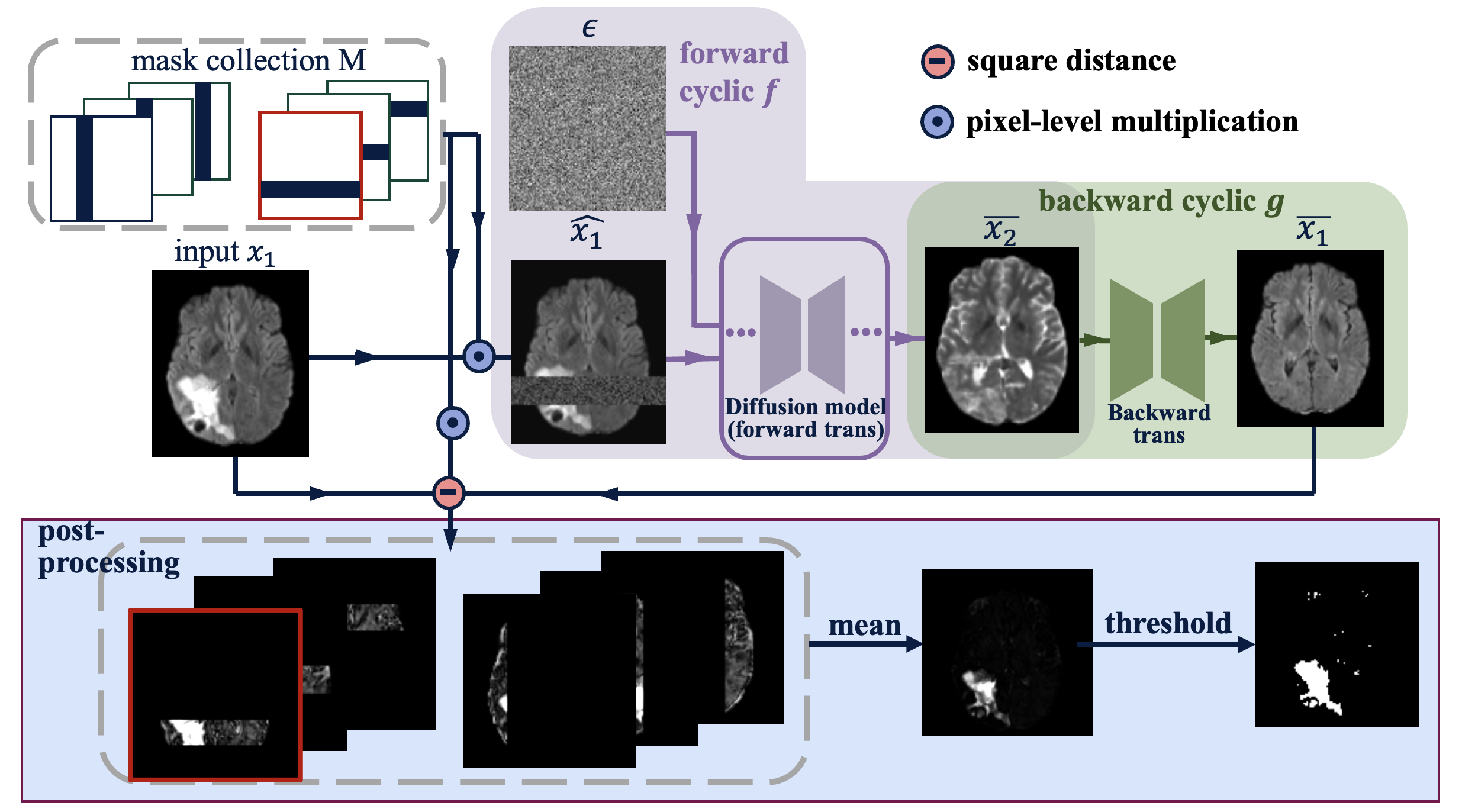}
\centering
\caption{The inference pipeline of our algorithm MMCCD. Image $\mathbf{x} \in \mathcal{X}$ is the input to the model. It is masked with $\mathbf{m_r}$ to $\mathbf{\hat{x}_{r}}$ and is the condition to the forward process, which is a conditional diffusion model. The forward process translates $\mathbf{x}$ to  $\mathbf{\overline{y}}$ from modality $\mathcal{Y}$. For backward translation, a UNet translates $\mathbf{\overline{y}}$ back to $\mathbf{\overline{x}}$. Finally, the anomaly is detected from the cyclic-translation error $\parallel \mathbf{x}-\mathbf{\overline{x}} \parallel _2$. We average the results from iterations using different masks and use a threshold $h$ to segment the anomaly.} \label{fig2}
\end{figure}

\noindent \textbf{Conditional Masked Diffusion Models: }
In a given input modality $\mathcal{X}$, it is possible that an unknown pattern encountered at test time, such as a tissue never seen during training, may exhibit an appearance that is not very distinctive in comparison to another `known' tissues. This may occur, for instance, when the input modality is not sensitive to an unknown encountered pathology, thus not highlighting it. In this case, the anomaly may be translated similar to a `known' tissue and back to the original modality with relatively low translation error, leading to suboptimal segmentation. In the experimental section, we later provide an example for this case (Fig.\ref{fig4}).

To alleviate this issue, a conditional generative model $f$ is employed to learn the translation $\mathcal{X} \rightarrow \mathcal{Y}$. We use conditional Denoising Diffusion Probabilistic Models (DDPMs)\cite{ho_denoising_2020} due to their capability of high-fidelity image generation. Basic diffusion models are unconditional and generate random in-distribution images. In recent works, various conditioning mechanisms have been proposed to guide the generation \cite{saharia_image_2021, dhariwal_diffusion_2021}. We use the input modality image as the condition for DDPM to realize translation, as described below. Furthermore, we iteratively mask different areas of the condition image $x\in \mathcal{X}$ with strong Gaussian noise and perform in-painting of the masked area. The model re-fills masked areas with generated in-distribution patterns, i.e. types of tissue encountered during training, that are consistent with the surrounding unmasked area. This aims to align the translated image $\overline{y}$ to the training distribution $\mathcal{Y}$. The assumption is that in-painting by DDPM will recreate normal tissues but not recreate the unknown patterns as the model does not model them, thus enlarging the translation error of the anomalous patterns, facilitating their segmentation.
We start by introducing the conditioned diffusion model used for translation, then we introduce the masked generation process.

Diffusion models have the \emph{forward} stage and the \emph{reverse} stage. In the \emph{forward} stage, denoted by $q$, the target image for the translation process $\mathbf{y_0}$ is perturbed by gradually adding Gaussian noise via a fixed Markov chain over $T$ iterations. The noisy image is denoted as $\mathbf{y_t}$ at iteration $t$. This Markov chain is defined by 
$q(\mathbf{y_{1:T}}\mid \mathbf{y_{0}})=\prod \limits_{t=1}^Tq(\mathbf{y_t}\mid \mathbf{y_{t-1}})$.
Each step of the diffusion model can be denoted as 
$q(\mathbf{y_t}\mid \mathbf{y_{t-1}})=\mathcal{N}(\mathbf{y_{t}}\mid\sqrt{\alpha_t}\mathbf{y_{t-1}},(1-\alpha_{t})\mathbf{I})$,
in which $0<\alpha_t<1$ follows a fixed schedule that determines the variance of each step's noise. Importantly,  we can characterize the distribution of $\mathbf{y_t}$ as 
\begin{equation}
\label{Equ:q_yt}
q(\mathbf{y_t}\mid \mathbf{y_0})=\mathcal{N}(\mathbf{y_t}\mid \sqrt{\overline{\alpha}_t}\mathbf{y_0}, (1-\overline{\alpha}_t)\mathbf{I}),
\end{equation} 
where $\overline{\alpha}_t = \prod \limits_{t=1}^t \alpha_i$. Furthermore, using the Bayes' theorem, we can calculate
\begin{equation}
\begin{aligned}
\label{Equ:q_y(t-1)}
&q(\mathbf{y_{t-1}}\mid \mathbf{y_t},\mathbf{y_0})=\mathcal{N}(\mathbf{y_{t-1}}\mid \mu_t, \sigma^2 \mathbf{I}),\\ 
&\textrm{in}\quad \textrm{which} \quad \mu_t=\frac{\sqrt{\overline{\alpha}_{t-1}}\beta_t}{1-\overline{\alpha}_t}\mathbf{y_0}+\frac{\sqrt{\alpha_t}(1-\overline{\alpha}_{t-1})}{1-\overline{\alpha}_t}\mathbf{y_t}, \\ 
&\textrm{and} \quad \sigma^2 = \frac{1-\overline{\alpha}_{t-1}}{1-\overline{\alpha}_t}(1-\alpha_t).
\end{aligned}
\end{equation}
The \emph{reverse} step, which is essentially the generative modeling, is also a Markov process denoted by $p$. This process aims to reverse the forward diffusion process step-by-step to recover the original image $y_0$ from the noisy image $\mathbf{y_T}$. In order to do this, the conditional generative model, given as condition an image $\mathbf{x}$ from modality $\mathcal{X}$, is optimized to approximate the posterior $q(\mathbf{y_{t-1}}\mid \mathbf{y_t})$ with 
\begin{equation}
p_\theta (\mathbf{y_{t-1}}\mid \mathbf{y_t}, \mathbf{x})=\mathcal{N}(\mathbf{y_{t-1}}\mid \mu_\theta(\mathbf{y_t}, \mathbf{x}, t), \sum_\theta(\mathbf{y_t}, \mathbf{x}, t)).
\end{equation} 
Note that in our setting, we have the image-level condition $x_0$ to guide the translation process. In this distribution, the variance $\sum_\theta(\mathbf{y_t}, \mathbf{x}, t)$ is fixed by the $\alpha_t$ schedule defined during the forward process. 
The input of the model is $\mathbf{y_t}$, whose marginal distribution is compatible with Eq.\ref{Equ:q_yt}: $\tilde{\mathbf{y_t}}=\sqrt{\overline{\alpha}}\mathbf{y_0}+(1-\overline{\alpha}_t)\epsilon$, where $\epsilon \sim \mathcal{N}(0,I)$. While we want to reverse $q$, the distribution becomes tractable when conditioned on $\mathbf{y_0}$ as in Eq.\ref{Equ:q_y(t-1)}. we parameterize the model to predict $ \mu_\theta(\mathbf{y_t}, \mathbf{x}, t)$ by predicting the original image $\mathbf{y_0}$. Therefore, the training objective of our diffusion model $f$ is:
\begin{equation}
L_f = \lVert f(\mathbf{y_t}, \mathbf{x}, \alpha_t) - \mathbf{y_0} \rVert _2
\end{equation}
Furthermore, to perform cyclic translation, we also train a deterministic translation model $g$ to translate the image from modality $\mathcal{Y}$ back to modality $\mathcal{X}$. 

\begin{equation}
L_g = \lVert g(\mathbf{y_0}) - \mathbf{x} \rVert _2
\end{equation}

Furthermore, we iteratively apply a set of masks to the condition image and use the generative model to recreate them according to the training data distribution. By applying a mask, the model can in-paint simultaneously with translation. During in-painting, we utilize the strong generative ability of the diffusion model to generate in-distribution patterns under the mask, i.e. types of tissue encountered during training, with the guidance of the surrounding unmasked area. Unknown, anomalous patterns are less likely to be recreated as the DDPM is not modeling them during training. Therefore the translation error after the cyclic process will be higher between the generated parts of the image and the real image when in-painting areas of anomalous patterns, than when in-painting `normal' (known) tissue types. Given a condition image $\mathbf{x}$ from the input modality, and a mask $\mathbf{m_r}$ where pixels with value $1$ indicate the area to be masked and $0$ for the unmasked area, the diffusion model becomes:
\begin{equation}
\label{Equ:mask_cond}
\mathbf{\hat{x}_{r}} = (1-\mathbf{m_r}) \odot \mathbf{x} + \mathbf{m_r} \odot \epsilon
\end{equation} 
where $\epsilon \sim \mathcal{N}(0,I)$, covering the masked area with white noise. During training, the mask is randomly positioned within the image. The training loss for the masked diffusion then becomes:
\begin{equation}
\label{Equ:final_loss}
L_f = \lVert f(\mathbf{y_t}, \mathbf{\hat{x}_r}, \alpha_t) - \mathbf{y_0} \rVert _2
\end{equation}
During testing, since we want to detect the abnormality in the whole image, we design a collection of masks, recorded as $\mathbf{M} = \left\{\mathbf{m_1}, \mathbf{m_2}, ..., \mathbf{m_R}\right\}$, where $R$ equals the total mask number. For each mask, the diffusion model $f$ performs the reverse steps $t=T, ...1$ using masked condition $\hat{x}_{r}$ (Eq.\ref{Equ:mask_cond}) to translate the condition $\hat{x}_r$ to corresponding $\overline{y}_0$ while in-painting masked areas. Then the output prediction $\overline{y}_0$ is given as input to backward translation model $g$ and is translated back to the original modality space $\mathcal{X}$ deterministically, without masking. The final abnormality is detected by:
\begin{equation}
\label{eq:pred_mmccd}
    \mathbf{a} = \frac{1}{\sum_{r=1}^{R}\mathbf{m_r}}\sum_{r=1}^{R}\mathbf{m_r}\lVert g(f(\mathbf{x}, \mathbf{m_r})) - \mathbf{x} \rVert _2 > h
\end{equation}
where the sums and division are performed voxel-wise, and $f$ denotes the complete diffusion process with $T$ diffusion step. The prediction from model $f(\mathbf{x}, \mathbf{m_r})$ is the input to $g$. The threshold for detecting anomalies is given by $h$. The pipeline for training and testing is shown in Alg.\ref{alg:training} and Alg.\ref{alg:testing} respectively.

\begin{minipage}{\textwidth}
  \centering
  \begin{minipage}[t]{.42\textwidth}
    \vspace{0pt}
    \centering
    \RestyleAlgo{ruled}
    \begin{algorithm}[H]
    \caption{Training}\label{alg:training}
    \KwData{$\!(\mathbf{x}, \mathbf{y}) \sim \left\{\mathbf{x^{i}},\mathbf{y^{i}}\right\}^N_{i=1}$}
    \BlankLine
    $\mathbf{m}_r \in \left\{\mathbf{m_1}, \mathbf{m_2}, ..., \mathbf{m_n}\right\}$\;
    $\epsilon_1, \epsilon_2, \epsilon_3, \epsilon_4 \in \mathcal{N}(0,1)$\;
    $\alpha_t \sim p(\alpha)$\;
    $\mathbf{\hat{y}_r} = (1-\mathbf{m_r}) \odot \mathbf{y} + \mathbf{m_r} \odot \epsilon_1$\;
    $\mathbf{\hat{x}_r} = (1-\mathbf{m_r}) \odot \mathbf{x} + \mathbf{m_r} \odot \epsilon_2$\;
    $\mathbf{\tilde{x_t}}=\sqrt{\overline{\alpha}_t}\mathbf{x}+(1-\overline{\alpha}_t)\epsilon_3$\;
    $\mathbf{\tilde{y_t}}=\sqrt{\overline{\alpha}_t}\mathbf{y}+(1-\overline{\alpha}_t)\epsilon_4$\;
    Gradient descent on:\ \ \ \ 
    $L_f\! = \lVert \! f(\mathbf{y_t}, \mathbf{\hat{x}_r}, \alpha_t) - \mathbf{y} \! \rVert _2$;
    \BlankLine
    $L_g = \lVert g(\mathbf{y}) - \mathbf{x} \rVert _2$;
    \end{algorithm}
    
  \end{minipage}
  \begin{minipage}[t]{.56\textwidth}
    \vspace{0pt}
    \centering
    \RestyleAlgo{ruled}
    \begin{algorithm}[H]
    \caption{Testing}\label{alg:testing}
    \KwData{$\mathbf{x} \sim \left\{\mathbf{x^{i}}\right\}^N_{i=1}$}
    $\mathbf{y_T} \sim \mathcal{N}(0,I), \mathbf{x_T} \sim \mathcal{N}(0,I)$\;
    \For{$\mathbf{m}_r \in \left\{\mathbf{m_1}, \mathbf{m_2}, ..., \mathbf{m_n}\right\}$}{
        $\epsilon_1 \in \mathcal{N}(0,1)$\;
        $\mathbf{\hat{x}_r} = (1-\mathbf{m_r}) \odot \mathbf{x} + \mathbf{m_r} \odot \epsilon_1$\; 
        \For{$t=T,...,1$}{
          $\epsilon \sim \mathcal{N}(0,I)$ if $t>1$, else $\epsilon=0$\;
          $\mathbf{\overline{y}_{t-1,r}}=\frac{\sqrt{\overline{\alpha}_{t-1}}\beta_t}{1-\overline{\alpha}_t}f(\mathbf{\hat{x}_r},\mathbf{\overline{y_{t,r}}},\alpha_t)+\frac{\sqrt{\alpha_t}(1-\overline{\alpha}_{t-1})}{1-\overline{\alpha}_t}\mathbf{\overline{y}_{t,r}}+\sqrt{1-\alpha_t}\epsilon$
        }
        $\mathbf{\overline{x}_{r}} = g(\mathbf{\overline{y}_{r}})$\;
    }
    $\mathbf{a} = \frac{1}{\sum_{r=1}^{n}\mathbf{m_r}}\sum_{r=1}^{n}\mathbf{m_r} \odot$
    $ \qquad \qquad \qquad \qquad \lVert \mathbf{\overline{x}_{r}} - \mathbf{x} \rVert _2 > h; $
    \end{algorithm}
  \end{minipage}
  \label{fig:twoalg}
\end{minipage}

\section{Experiments}
\noindent \textbf{Dataset:}
For evaluation, as commonly done in literature for unsupervised anomaly segmentation, we evaluate our method on the proxy task of learning normal brain tissues during training and attempt to segment brain pathologies in an unsupervised manner at test time. To conduct the evaluation, we use the BraTS2021 dataset\cite{bakas_advancing_2017, baid_rsna-asnr-miccai_2021}. 80\% of the images are used for training, 10\% for validation, and 10\% for testing. We used FLAIR, T1, and T2 in our various experiments. We normalized each image by subtracting the mean and dividing by the standard deviation of the brain intensities between the 2\% and 98\% percentiles for T1, T2, and 2\%-90\% for FLAIR which presents more extreme hyper-intensities.
We evaluate our method using 2D models, as commonly done in previous works to simplify experimentation. For this purpose, from every 3D image, we extract slices 70 to 90, that mostly capture the central part of the brain. For model training, from the slices extracted from the training subjects, we only use those that do not contain any tumors. For validation and testing of all compared methods, from each validation and test subject, we use the slice that contains the largest tumor area out of the 20 central slices.

\noindent \textbf{Model Configuration:} The diffusion model, Cyclic UNet, and MMCCD use the same UNet as in \cite{dhariwal_diffusion_2021} for a fair comparison. The image condition is given by concatenating $\mathbf{x}$ as an input channel with $\mathbf{y_t}$ as in\cite{saharia_image_2021}. Adam optimizer is used and the learning rate is set to $1e-4$. The model is trained with a batch size of 32. To accelerate the sampling process of the diffusion model, the method of DDIM\cite{song_denoising_2022} is used that speeds up sampling tenfold. Input slices are resampled to $128 \times 128$ pixels to be compatible with this UNet architecture. For MMCCD, the mask's size is $16\times128$ pixels. To ensure complete coverage of the brain, we gradually move the mask with a stride of $2$, obtaining one $\mathbf{m_r}$ (Eq.\ref{eq:pred_mmccd}) for each valid position, along with the corresponding prediction. This is repeated for horizontal and vertical masks. All hyper-parameters for our method and compared baselines were found on the validation set.

\noindent \textbf{Evaluation:}
We compare a variety of unsupervised abnormality segmentation methods. We assess as baselines an AE and DAE \cite{kascenas_denoising_2022}, VAE\cite{baur_autoencoders_2021}, and DDPM\cite{pinaya_fast_2022}. In the first setting, as common in previous works, we compare methods for unsupervised segmentation of tumors in the FLAIR modality. We train the aforementioned models on normal-looking FLAIR slices. We then test them on FLAIR slices from the test set that includes tumor. To show the potential of cyclic modality translation for unsupervised anomaly segmentation, we assess a Cyclic UNet, where $f$ and $g$ are two UNets trained separately. In this setting, the two UNets are trained on normal-looking slices where $\mathcal{X}$ is FLAIR and $\mathcal{Y}$ is T2 modality. We emphasize that the real FLAIR and T2 images are only needed during training, whereas the model only needs FLAIR during testing, making Cyclic UNet comparable with the aforementioned baselines at test time. Similarly, to assess the added effect of masking, we train MMCCD similarly to Cyclic UNet. We show performance on the test set in Tab.\ref{tab1}. Performance is evaluated using Dice coefficient (DICE), area under the curve (AUC), Jaccard index (Jac), precision (Prec), recall (Rec), and average symmetric surface distance (ASSD).

AE and VAE baselines show what a basic reconstruction model can achieve. DAE further improves the abnormality segmentation performance by introducing Gaussian noise.
It is likely that more complex noise may improve results, but the type of optimal noise may be dependent on the specific studied task of anomaly segmentation. Since the effect of noise is not studied in this work, we only include the results using Gaussian noise. This also facilitates fair comparison with the DDPM model, which also uses Gaussian noise for distorting the image, similar to \cite{pinaya_fast_2022}. DDPM outperforms the AE, VAE, and DAE baselines. We did not include GAN-based approaches as the DDPM-based model is reported to outperform GAN-based approaches in other works \cite{pinaya_fast_2022,kascenas_denoising_2022} and it is notoriously challenging to train adequately.
The high performance of Cyclic UNet is evidence that relying on the mapping between modalities is a very promising mechanism for unsupervised anomaly segmentation. The simplest Cyclic UNet, without any noise or masking, significantly outperforms methods relying on reconstructing (i.e. approximate the identity function) in most metrics with AE and VAE, or denoising with DAE and DDPM. Combining the cyclic translation with a diffusion model and masking further improves performance, achieving the best results in almost all metrics. Finally, we note that masking can be viewed as a type of noise, and thus we speculate that the proposed cyclic-translation mechanism could be combined effectively with other types of noise and not just the proposed masking mechanism, but such investigation is left for future work.

\begin{table}[]
\centering
\caption{Performance for unsupervised anomaly segmentation of tumors with FLAIR as input modality $\mathcal{X}$. Cyclic UNet and MMCCD translate FLAIR ($\mathcal{X}$) to T2 ($\mathcal{Y}$) and back. Best results are shown in bold.}
\label{tab1}
\begin{tabular}{>{\centering\arraybackslash}p{22mm}>{\centering\arraybackslash}p{15mm}>{\centering\arraybackslash}p{15mm}>{\centering\arraybackslash}p{15mm}>{\centering\arraybackslash}p{15mm}>{\centering\arraybackslash}p{15mm}>{\centering\arraybackslash}p{15mm}}
\hline
            & \textbf{DICE}   & \textbf{AUC}    & \textbf{Jac}    & \textbf{Prec}   & \textbf{Rec} & \textbf{ASSD}   \\ \hline
\textbf{AE\cite{baur_autoencoders_2021}}         & 0.2249 & 0.4621 & 0.1513 & 0.2190 & 0.2636 & 3.5936  \\
\textbf{VAE\cite{baur_autoencoders_2021}}         & 0.2640 & 0.4739 & 0.1910 & 0.2717 & 0.2837 & \textbf{3.513}  \\
\textbf{DAE\cite{kascenas_denoising_2022}}         & 0.4619 & 0.9298 & 0.3291 & 0.4346 & 0.5477 & 7.5508 \\
\textbf{DDPM\cite{pinaya_fast_2022}}        & 0.5662 & 0.9267 & 0.4172 & 0.5958 & 0.5949 & 6.5021 \\
\hline
\textbf{Cyclic UNet} & 0.5810 & 0.9409 & 0.4470 & \textbf{0.6545} & 0.5823 & 6.3487 \\
\textbf{MMCCD}        & \textbf{0.6092} & \textbf{0.9409} & \textbf{0.4682} & 0.6505 & \textbf{0.6336} & 6.1171 \\ \hline
\end{tabular}
\end{table}

\begin{figure}[htbp]
\centering
\includegraphics[width=0.8\textwidth]{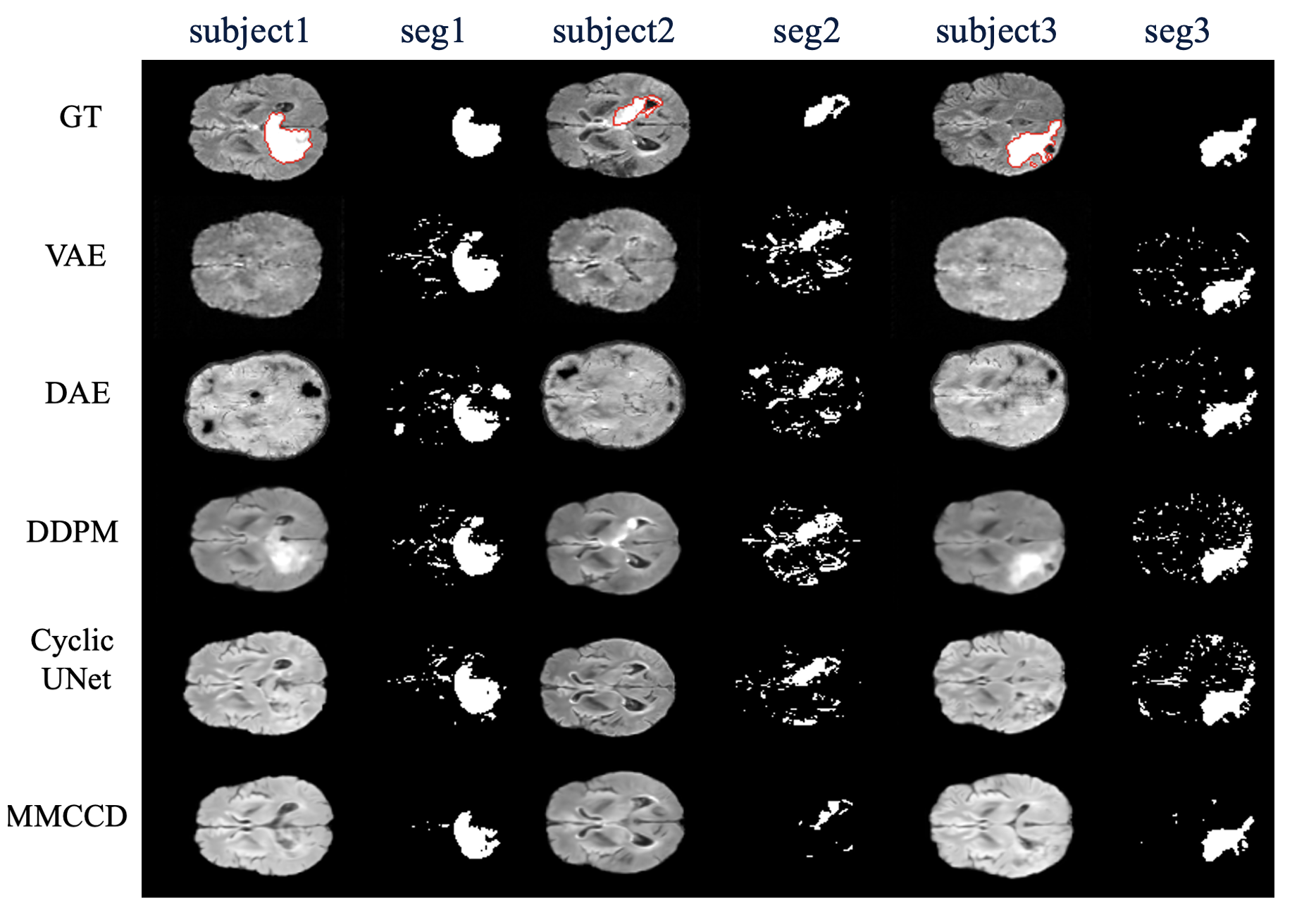}
\caption{The visualization results for different methods on three examples from FLAIR modality. Each row from top to bottom: ground-truth of the image/ segmentation label, VAE\cite{baur_autoencoders_2021}, DAE\cite{kascenas_denoising_2022}, DDPM\cite{pinaya_fast_2022}, Cyclic UNet, and MMCCD.} \label{fig3}
\end{figure}

We further evaluate the methods using different MRI modalities for $\mathcal{X}$ and $\mathcal{Y}$, as shown in Table \ref{tab2}. In these experiments, input modality $\mathcal{X}$ refers to the modality in which we aim to segment the anomalous patterns. Modality $\mathcal{Y}$ serves as the target for translation by Cyclic Unet and MMCCD. We compare with DAE, as it is the auto-encoder model found most promising in experiments of Table \ref{tab1}, and DDPM for comparing with the state-of-the-art diffusion-based method. For DAE and DDPM, only the first modality is used for both training and testing. For Cyclic UNet and MMCCD, which include the cyclic translation process, we include settings on FL-T2-FL, FL-T1-FL, T2-FL-T2, T2-T1-T2, T1-T2-T1, and T1-FL-T1. During testing, Cyclic UNet and MMCCD are tested using only inputs of modality $\mathcal{X}$. Table \ref{tab2} reports results using DICE.
These experiments show that for all methods,  as expected, the visibility and distinctiveness of tumors in input $\mathcal{X}$ largely affects segmentation quality. Using FLAIR as input yields the highest scores because tumors in FLAIR exhibit a distinctive intensity response, deviating significantly from the normal tissue. Using T1 as input results in the lowest performance because only part of the tumor is visible in this modality. 
\begin{table}
\centering
\caption{Performance for different combination of $\mathcal{X}$ and $\mathcal{Y}$ modalities of BraTS. DICE is reported. (FL is FLAIR)}\label{tab2}
\begin{tabular}{>{\centering\arraybackslash}p{40mm} >{\centering\arraybackslash}p{12mm}>{\centering\arraybackslash}p{12mm}>{\centering\arraybackslash}p{12mm}>{\centering\arraybackslash}p{12mm}>{\centering\arraybackslash}p{12mm}>{\centering\arraybackslash}p{12mm}}
\hline
\textbf{Input modality ($\mathcal{X}$)}  & \textbf{FL}     & \textbf{FL}     & \textbf{T2}     & \textbf{T2}     & \textbf{T1}     & \textbf{T1}     \\
\textbf{Translated modality ($\mathcal{Y}$)} & \textbf{T2}     & \textbf{T1}     & \textbf{FL}     & \textbf{T1}     & \textbf{T2}     & \textbf{FL}     \\ \hline
\textbf{DAE\cite{kascenas_denoising_2022}}  & 0.4619 & 0.4619 & 0.2803 & 0.2803 & 0.2325 & 0.2325 \\
\textbf{DDPM\cite{pinaya_fast_2022}}            & 0.5662 & 0.5662 & 0.4633 & 0.4633 & 0.2865 & 0.2865 \\
\hline
\textbf{Cyclic UNet}     & 0.5810 & 0.5732 & 0.4819 & 0.4036 & 0.3262 & 0.3906 \\
    \textbf{MMCCD}            & \textbf{0.6092} & \textbf{0.6090} & \textbf{0.4873} & \textbf{0.4993} & \textbf{0.3337} & \textbf{0.4239} \\ \hline
\end{tabular}
\end{table}

\begin{figure}
\centering
\includegraphics[width=0.5\textwidth]{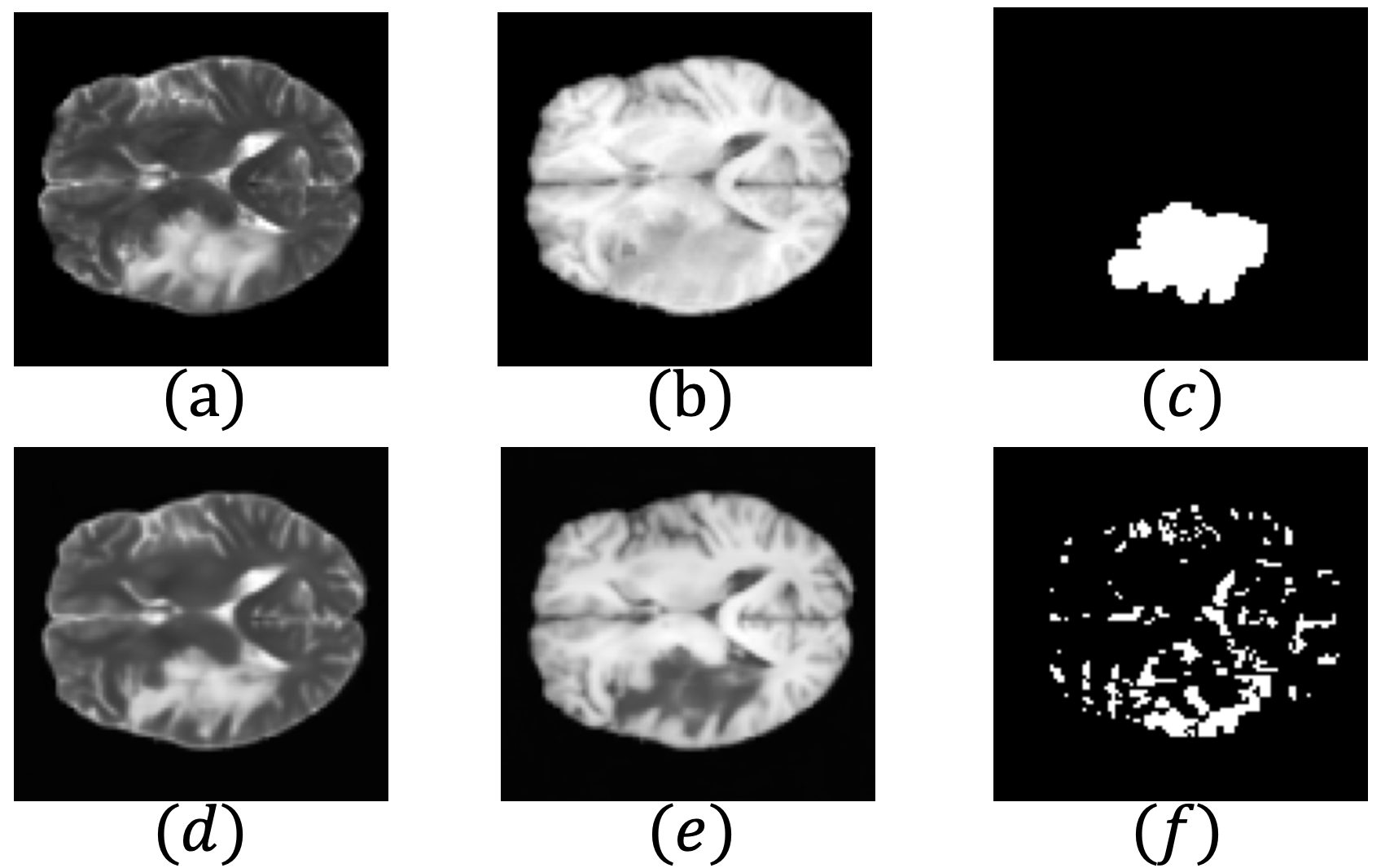}
\caption{Example failure case of cyclic translation T2-T1-T2 with Cyclic Unet, because anomaly exhibits similar appearance to normal tissues in the input modality. (a) input modality T2, (b) target modality T1, (c) tumor ground-truth, (d) cyclic translation result $\overline{x}$, (e) forward translation result $\overline{y}$, (f) predicted anomaly.}
\label{fig4}
\end{figure}

As Table \ref{tab2} shows, Cyclic UNet outperforms compared methods in most settings, supporting that cyclic translation is a promising approach regardless the modality of the data. One exception arises in the T2-T1-T2 setting, and we attribute this to the similar response to the RF signal in T1 from tumor tissues and non-tumor tissues, as shown in Fig.\ref{fig4}. In this example, the tumor area is erroneously detected as a ventricle and was recreated after cyclic translation, leading to low cyclic-translation error under the tumor area. In this setting, MMCCD that integrates masking and diffusion models exhibits an improvement of 9.57\% over Cyclic UNet, and outperforms all methods, highlighting the complementary effectiveness of generative in-painting with modality-cycles.

When input modality $\mathcal{X}$ is T1, the tumor's intensity falls within the range of normal tissue, and it's easy for the network to reconstruct the tumor well instead of perceiving it as unknown, making it challenging to detect anomalies. In this setting, the Cyclic UNet shows noticeable improvements, with 3.97\% improvement on T1-T2-T1 setting and a 10.41\% on T1-FL-T1, demonstrating Cyclic UNet's ability to effectively capture the subtle differences between tissues.

\section{Conclusion}
This paper proposed the use of cyclic translation between modalities as a mechanism for unsupervised abnormality segmentation for multimodal MRI. Our experiments showed that translating between modalities with a Cyclic UNet outperforms reconstruction-based approaches with autoencoders and state-of-the-art denoising-based approaches with diffusion models. We further extend the approach using a masked conditional diffusion model to incorporate in-painting into the translation process. We show that the resulting MMCCD model outperforms all compared approaches for unsupervised segmentation of tumors in brain MRI. The method requires multi-modal data for training, but only one modality at inference time. These results demonstrate the potential of multi-modal translation as a mechanism for facilitating unsupervised anomaly segmentation.

\subsubsection{Acknowledgements.}
ZL and HA are supported by scholarships provided by the EPSRC Doctoral Training Partnerships programme [EP/W524311/1]. FW is supported by the EPSRC Centre for Doctoral Training in Health Data Science (EP/S02428X/1), by the Anglo-Austrian Society, and by the Reuben Foundation.
The authors also acknowledge the use of the University of Oxford Advanced Research Computing (ARC) facility in carrying out this work(http://dx.doi.org/

\noindent 10.5281/zenodo.22558). 

\bibliographystyle{spIncs04_with_et_al}

\bibliography{references}

\end{document}